\documentstyle[epsfig,aps,prb,amsmath]{revtex}
\newcommand\beq{\begin{equation}}
\newcommand\eeq{\end{equation}}
\newcommand\bea{\begin{eqnarray}}
\newcommand\eea{\end{eqnarray}}

\begin{document}

\begin{center}
{\Large Study of Low-energy States of Clusters of Spin-1/2 and Spin-1 
Triangles with Kagome-like Geometries}
\end{center}

\vskip .5 true cm
\centerline{\bf Indranil Rudra$^1$, Diptiman Sen$^2$ and S. Ramasesha$^1$} 
\vskip .5 true cm

\centerline{\it $^1$ Solid State and Structural Chemistry Unit}
\centerline{\it $^2$ Centre for Theoretical Studies}
\centerline{\it Indian Institute of Science, Bangalore 560012, India}
\vskip .5 true cm

\begin{abstract}

We study the low-energy properties of Heisenberg antiferromagnetic spin-1/2 and
spin-1 systems on various clusters made up of triangles. Some of the clusters 
have a geometry similar to representative pieces of the Kagome lattice,
while others have the geometry of a sawtooth chain. While the ground state
always has the lowest possible spin (0 or 1/2), the nature of the low-energy
excitations depends on the geometry and the site spin. For the Kagome clusters
with spin-1 sites, the lowest excitations are gapped, with singlet and triplet
excitations having similar gaps. This is in contrast to Kagome clusters with 
spin-1/2 sites where there are several low-energy singlet excitations (possibly
gapless in the thermodynamic limit), while triplet excitations have a gap.
For the sawtooth chain with spin-1 sites, the lowest excitations are triplets 
with a gap; the gap to singlet excitations is about twice the triplet gap. 

\end{abstract}
\vskip .5 true cm

~~~~~~ PACS number: ~75.10.Jm, ~75.50.Ee

\newpage

There has been a great deal of interest in recent years in quantum spin systems
which are strongly frustrated \cite{lhuillier}. Examples of such systems 
include lattices involving triangles of Heisenberg spins as motifs which 
interact antiferromagnetically with each other. Some of the systems which
have been studied experimentally in one and two dimensions are the sawtooth 
chain \cite{saw} and the Kagome lattice \cite{kag1}. A variety of theoretical 
techniques, both analytical and numerical, have been used to study models of 
such systems \cite{kag2,kag3,lech,sind,kagstrip,hida,mila,raghu}. 

Classically (i.e., in the limit in which the site spin $S \rightarrow \infty$),
the sawtooth and Kagome antiferromagnetic
systems have an enormous ground state degeneracy arising from local 
rotational degrees of freedom which cost no energy; this leads to an 
extensive entropy at zero temperature. Quantum mechanically, this degeneracy 
is lifted due to nonzero matrix elements between different classical ground 
states. However, one might still expect a remnant of the classical degeneracy 
in the form of a large number of low-energy excitations. In the quantum limit,
it is well-known that the physics of some spin-1/2 and spin-1 lattices in low 
dimensions is different in nontrivial ways. It is therefore useful to study 
the lattices involving triangular motifs for site spin-1/2 and spin-1 to see 
if there is any systematic difference between the two cases.

The spin-1/2 Kagome system is known to exhibit a variety of unusual low-energy
properties \cite{lech,sind}. For instance, it is believed that there is a 
small gap between the singlet ground state and the nonsinglet excited states. 
Secondly, there is a band of singlet excited states with no band gap to the 
ground state. Finally, the two-spin correlation 
function decays rapidly with distance indicating that the magnetic correlation
length is comparable to the lattice spacing. All these features can be 
explained qualitatively by assuming that a set of disjoint triangles are in 
their ground states with total spin-1/2. Then these spin-1/2 states of nearest 
neighbor triangles can form dimers, thus producing a singlet state \cite{mila}.
The number of such dimer configurations is about $1.15^N$ if the number of 
sites is $N$; this number is macroscopically large even though it is much 
smaller than the number of singlets which one can form out of $N$ spin-1/2 
objects which is about $2^N$. The gapless band of singlet excitations can be 
thought of as arising from these dimer states, with a small dispersion being 
produced by the admixture amongst the different dimer states. In contrast to 
the spin-1/2 case, the ground state of a single spin-1 triangle is unique with
a gap to all its excited states. Following arguments similar to the one given 
above for the spin-1/2 Kagome lattice, one might then expect the ground state 
of the spin-1 Kagome lattice to be unique with a gap to {\it all} excitations.

For the spin-1/2 sawtooth chain, the ground state has a degeneracy proportional
to the number of sites. This degeneracy arises from the existence of spin-1/2
kinks (a free spin located at the base of a triangle shown in Fig. 6)
which do not cost any energy regardless of their lattice position
\cite{saw}. There are also spin-1/2 antikinks (a free spin located at the top
of a triangle in Fig. 6) which cost a finite energy. The lowest excitations 
in a chain with periodic boundary conditions are given by a kink-antikink 
pair which has a gap; the total spin of the pair may be either 0 or 1. 
The spin-1 sawtooth chain had not been studied earlier. We will present a 
numerical study of this system and will also construct an approximate
analytical ground state using a simple physical picture.

In this paper, we will carry out exact diagonalization studies of the 
Hamiltonian for both spin-1/2 and spin-1 Kagome clusters as well as the spin-1
sawtooth chain. We will discuss the numerical results for the low-energy 
spectrum in the different cases, and will use trial wave functions to 
provide an understanding of the results for the spin-1 sawtooth chain. 

We will study the lattices of spins shown in Figs. 1-6. Figs. 1-5 
show Kagome-type clusters, with the largest number of sites being 21 for
spin-1 and 27 for spin-1/2. Fig. 6 shows a 21-site sawtooth lattice which we 
have studied for site spin-1. The Hamiltonian used in our calculation is the 
Heisenberg Hamiltonian with isotropic exchange, written as
\beq
{\hat H} ~=~ J~ \underset{<ij>}{\Sigma} ~{\hat {\bf S}}_i \cdot {\hat 
{\bf S}}_j ~,
\label{ham}
\eeq
where $J$ is the exchange interaction, and $<ij>$ sums over 
the bonded sites shown in Figs. 1-6.
We have imposed periodic boundary conditions at the edges of the various 
clusters as indicated by the dotted lines in Figs. 1-6. 
The properties of the system can be computed from the eigenstates of
the Hamiltonian which are obtained by setting up the Hamiltonian
matrix in a suitable basis and diagonalizing it thereafter. The dimension of
the space spanned by the Hamiltonian rapidly increases with the number of 
spins in the system. Though the use of a constant $M_S$ (i.e., eigenstates of 
total $S^z$) basis is quite straightforward, the dimensions of the space for 
large clusters which have no spatial symmetry are prohibitively large. Hence 
we have used the valence bond (VB) technique \cite{ramasesha} to construct 
spin adapted functions (eigenstates of the total ${\hat {\bf S}}^2$ operator).
The advantage of this method is that the dimensions of the constant total $S$ 
subspaces are much smaller compared to the constant $M_S$ subspaces. 
After constructing the VB basis, we obtain the Hamiltonian matrix by operating
with Eq. (\ref{ham}) on each of the basis states. The matrix representing the 
Hamiltonian in the VB basis is in general nonsymmetric since the VB basis is 
nonorthogonal. We use Rettrup's algorithm \cite{ret} to obtain the lowest few 
eigenstates and eigenvectors in each of the total S subspaces. We have 
transformed the nonorthogonal VB basis to the orthogonal constant $M_S$ basis
which are also eigenstates of the $z$ component of the site spins. This
enables us to calculate the spin-spin correlation functions, $<S_i^z S_j^z>$,
for the different clusters. 

For the spin-1 Kagome clusters, we note that the ground state is always a 
singlet. The lowest excited states are gapped, with the gap for singlet
and triplet excitations being approximately equal as shown in Fig. 7. 
There is a good trial wave function for the singlet ground 
state due to Hida \cite{hida}. We consider each spin-1 as being made out of 
two spin-1/2's in a symmetrized combination. For
the Kagome clusters that we have considered, we see that each spin-1 site
belongs to only two hexagons. We can therefore associate each of the composite
spin-1/2's (forming a spin-1) with one hexagon. For each hexagon, 
we then consider the state $\psi_0$ which is the singlet ground state that 
arises when six spin-1/2's interact antiferromagnetically with nearest 
neighbors. A trial wave function for the ground state of the whole 
cluster is then given by a product of $\psi_0$'s over all the hexagons, which 
is then symmetrized between all the pairs of composite spin-1/2's.
This trial wave function gives a good estimate of the ground state 
energy \cite{hida}. However, this approximate picture fails to provide a good 
understanding of the spectrum of low-energy excited states. For six
spin-1/2's which form a hexagon and interact antiferromagnetically, the
lowest excited state is a triplet and the first singlet excitation lies
above this triplet state by about 1.3 $J$. However, for the spin-1 Kagome
clusters, we find numerically that the singlet and triplet excitations have 
comparable gaps to the ground state; excitations with spin-2 have a much 
higher energy. It would be very useful to understand these features of spin-1 
Kagome systems, even qualitatively through variational wave functions.

For the spin-1/2 Kagome clusters, we see from Fig. 8 that there is a 
large number of low-lying states which have the same spin as the ground state.
It is possible that in the thermodynamic limit, these states will form a 
band with the band bottom touching the ground state. On the other hand, the 
first excited state with spin one higher than the ground state spin seems to
be separated from the ground state by a finite gap. All this is in accordance
with the numerical results obtained earlier in Ref. [6], and can be understood
from the exponentially large number of dimer configurations 
as mentioned above \cite{mila}. 

We now turn to the two-spin correlation functions $< S_i^z S_j^z >$, where
$i,j$ denote two sites of a Kagome cluster. Table 1 shows that the correlations
fall off very rapidly beyond the second neighbor for the spin-1 case (with 21 
sites as in Fig. 3) and beyond the first neighbor for the spin-1/2 case (with 
27 sites as in Fig. 5). Thus the magnetic correlation length is very short, 
although it appears to be a bit longer for the spin-1 Kagome clusters
compared to spin-1/2. However, within the second neighbor sites 
for spin-1 and within the first neighbor sites for spin-1/2, there is no 
obvious pattern in the correlation functions. For instance, within the 
triangle 1-2-3 in the spin-1 case (Fig. 3) and the triangle 13-14-15 in the 
spin-1/2 case (Fig. 5), the three nearest neighbor correlations are not equal.
Similarly, in the spin-1 case, the correlation between the second 
neighbor sites 2-4 is about twice as big as the correlation between the
first neighbor sites 3-4. It is possible that these peculiarities are 
due to finite size effects, which would indicate an unusual sensitivity of 
the correlation functions to the boundary conditions. It would be very useful 
to understand these patterns of correlation functions even qualitatively.

Finally, we come to the the spin-1 sawtooth chain (Fig. 6). The energies of 
the low-lying are shown in Fig. 9. We see that there is a gap to the triplet 
excitation; the singlet excitations have a gap which is about twice as large 
as the triplet excitation gap. This clearly suggests that the elementary 
excitations of the spin-1 sawtooth chain are triplets, and the lowest singlet 
excitations are probably made up of two well-separated excited triplets.

There is an approximate way of visualizing the ground state and triplet
excitations of the spin-1 sawtooth chain. Let us first introduce an algebraic
parametrization for spins \cite{arovas}. Let $u = \cos \theta e^{i\alpha}$ and
$v = \sin \theta e^{i\beta}$ denote the spin-$\uparrow$ and spin-$\downarrow$
states respectively of a spin-1/2 object. The angles $\theta , \alpha$ and 
$\beta$ parametrize the surface of the sphere $S^3$, lying
in the ranges $0 \le \theta \le \pi /2$, $0 \le \alpha , \beta \le 2 \pi$. 
Then the three states of a spin-1 object are given by 
\bea
| S^z = 1 > ~&=&~ {\sqrt 3} ~u^2 ~, \nonumber \\
| S^z = 0 > ~&=&~ {\sqrt 6} ~uv ~, \nonumber \\
| S^z = -1 > ~&=&~ {\sqrt 3} ~v^2 ~, 
\eea
where the normalization factors on the right hand sides are found by 
integrating with the measure $d\Omega = d\theta d\alpha d\beta \sin \theta 
\cos \theta /(2\pi^2)$. (For instance, $\int d\Omega u^{*2} u^2 = 1/3$). Given
two spin-1/2 objects at sites $i$ and $j$, the singlet state $\psi_{i,j}$ is 
given by $(u_i v_j - u_j v_i)$. Then a trial ground state for the
(infinitely long) spin-1 sawtooth lattice is given by
\beq
\Psi_0 ~=~ \prod_{i=-\infty}^{\infty} ~\psi_{i,i+1} ~.
\label{trial}
\eeq
This is the AKLT state involving all the sites of the chain \cite{aklt}. In 
this state, the expectation values of $<{\hat {\bf S}}_i \cdot 
{\hat {\bf S}}_{i+1} >$ and $<{\hat {\bf S}}_i \cdot {\hat {\bf S}}_{i+2} >$ 
are equal to $-4/3$ and $4/9$ respectively \cite{arovas}. For the sawtooth 
chain, this trial state gives an energy per site, $e_0$, equal to $- 10/9 J 
\simeq -1.111 J$, compared to the numerically obtained value of $-1.117 J$. 
The trial state in Eq. (\ref{trial}) is the unique state which has the maximum
number of singlets between the composite spin-1/2 objects (of the site 
spin-1's) on nearest neighbor sites. It is therefore not surprising that it 
gives such a good estimate of the ground state energy.

A trial state for the lowest triplet state is given by a superposition of 
the form
\beq
\Psi (k) ~=~ \sum_n ~e^{ikn} ~(\prod_{i=-\infty}^{2n-2} ~\psi_{i,i+1}) ~
\psi_{2n-1,2n+1} ~(\prod_{i=2n+1}^{\infty} ~\psi_{i,i+1}) ~,
\label{triplet}
\eeq
where $k$ denotes the momentum of the state. Each of the wave functions on the
right hand side (RHS) of Eq. (\ref{triplet}) have the form of a AKLT state 
which omits the site labeled $2n$ (which lies at the top of a triangle in Fig.
6); it therefore describes a state with total spin-1. It is difficult to 
compute the expectation value of the Hamiltonian in the state in Eq. 
(\ref{triplet}), particularly because the different states on the RHS are not 
orthogonal. However, we expect this state to have the lowest energy amongst 
the triplet trial states because each of the states on the RHS of Eq. 
(\ref{triplet}) is the unique (up to translations) total spin-1 state having 
the maximum number of singlets between the composite spin-1/2's on nearest 
neighbor sites. Similarly, a trial state for the next excited state can be 
constructed as a superposition of AKLT states which omit two spin-1's, say, at
sites $2m$ and $2n$ and with wave numbers $k_1$ and $k_2$. The total spin of 
such states can be either $0$, $1$ or $2$. The bottom of this band may be 
expected to correspond to a gap which is about twice as large as the gap to the
bottom of the band with a single "free" spin-1 excitation. This appears
to be the case for the low-energy spectrum of the 19-site spin-1 sawtooth chain
shown in Fig. 9; we see that the lowest spin-0 and spin-2 excitations 
have a gap which is about twice as large as the gap to triplet excitations.
For the 21-site chain, we could only get three singlets and two triplets
since the Hilbert space of the system is much larger than for the 19-site 
system. The low-energy spectrum for the states common to the 19-site and
21-site systems are very similar.

To summarize, we have numerically studied a variety of spin-1 and spin-1/2
clusters with triangular motifs. We find significant 
differences between the low-energy spectra of the spin-1/2 and spin-1 systems.
For the Kagome clusters, the spin-1/2 model has a large number of possibly
gapless singlet excitations and a finite gap to the lowest triplet 
excitation. Hence the specific heat goes to zero as a power law ($T^2$), 
while the magnetic susceptibility falls off exponentially
as $\exp (-\Delta E_T / k_B T)$ on approaching 0 $K$ \cite{sind}. On the 
other hand, the spin-1 Kagome clusters have a gap to all 
excitations; hence both the specific heat and the magnetic susceptibility
are expected to fall off exponentially as one approaches 0 $K$. It is 
difficult to conclude from our studies if these differences are generic to 
integer spin versus half-odd-integer spin Kagome lattices. For the sawtooth 
chain, the spin-1/2 model is known to have a number of degenerate ground 
states; the lowest excitations are singlets and triplets with the same gap 
\cite{saw}. The spin-1 sawtooth chain has a nondegenerate ground state; the 
lowest excited states are triplets which have a nonzero gap to the ground 
state. We have presented a simple physical picture of the low-energy
states of the spin-1 sawtooth chain. Our numerical results will be useful
for developing a quantitative analytical understanding 
of the low-energy states of the spin-1 Kagome clusters.

We thank C. Lhuillier for drawing our attention to the spin-1 clusters. DS 
thanks DST, India for financial support under project SP/S2/M-11/00. SR 
thanks CSIR, India (CSIR/SR/CSS/187) and DAE-BRNS, India (DAE0/SRS/CSS/091) 
for financial support.

\newpage

\begin{center}
{\Large {\bf Spin-spin correlation functions for a 21 site spin-1 cluster}} 
\vspace{0.5cm}

\begin{tabular}{|c|c|c|c|}
\hline {\bf Site i} & {\bf Site j} & {\bf S=0 Ground State} & {\bf S=1 
Excited State} \\ \hline

1 & 1 & 0.667 & 0.656 \\ \hline
2 & 2 & 0.667 & 0.793 \\ \hline
3 & 3 & 0.667 & 0.656 \\ \hline
1 & 2 & -0.171~ & -0.222~ \\ \hline
1 & 3 & -0.296~ & -0.247~ \\ \hline
2 & 3 & -0.171~ & -0.222~ \\ \hline
1 & 4 & -0.021~ & 0.006 \\ \hline
1 & 5 & 0.076 & -0.007~ \\ \hline
1 & 6 & -0.020~ & -0.014~ \\ \hline
2 & 4 & -0.273~ & -0.180~ \\ \hline
2 & 5 & 0.055 & 0.150 \\ \hline
2 & 6 & 0.000 & 0.030 \\ \hline
3 & 4 & 0.138 & 0.078 \\ \hline
3 & 5 & -0.136~ & -0.099~ \\ \hline
3 & 6 & 0.018 & -0.005~ \\ \hline

\end{tabular}
\end{center}

\begin{center}
{\Large {\bf Spin-spin correlation functions for a 27 site spin-1/2 cluster}} 
\vspace{0.5cm}

\begin{tabular}{|c|c|c|c|}
\hline {\bf Site i} & {\bf Site j} & {\bf S=0.5 Ground State} & {\bf S=1.5 
Excited State} \\ \hline
13 & 13 & 0.250 & 0.250 \\ \hline
14 & 14 & 0.250 & 0.250 \\ \hline
15 & 15 & 0.250 & 0.250 \\ \hline
13 & 14 & -0.128~ & -0.050~ \\ \hline
14 & 15 & 0.025 & -0.107~ \\ \hline
13 & 15 & -0.128~ & -0.050~ \\ \hline
14 & 10 & -0.010~ & -0.009~ \\ \hline
14 & 11 & 0.003 & 0.003 \\ \hline
14 & 12 & 0.002 & 0.001 \\ \hline
13 & 10 & 0.002 & 0.008 \\ \hline
13 & 11 & 0.004 & 0.020 \\ \hline
13 & 12 & -0.009~ & -0.009~ \\ \hline
15 & 10 & 0.004 & 0.004 \\ \hline
15 & 11 & -0.012~ & -0.013~ \\ \hline
15 & 12 & 0.005 & 0.001 \\ \hline

\end{tabular}
\end{center}

Table 1. Spin-spin correlation functions $<S_i^z S_j^z>$ for the 21 site 
spin-1 Kagome cluster and the 27 site spin-1/2 Kagome cluster.

\newpage

\noindent {\bf Figure Captions}
\vskip .5 true cm

\noindent {1.} A Kagome cluster of 15 sites.

\noindent {2.} A Kagome cluster of 18 sites.

\noindent {3.} A Kagome cluster of 21 sites.

\noindent {4.} A Kagome cluster of 24 sites.

\noindent {5.} A Kagome cluster of 27 sites.

\noindent {6.} A sawtooth chain of 21 sites.

\noindent {7.} Low-lying energy levels of three spin-1 Kagome clusters (in
units of $J$). The singlet gaps are equal to 0.28 $J$, 0.27 $J$ and 0.25 $J$. 
The triplet gaps are equal to 0.35 $J$, 0.31 $J$ and 0.28 $J$.

\noindent {8.} Low-lying energy levels of three spin-1/2 Kagome clusters (in
units of $J$). The lowest doublet/singlet gaps are equal to 0.04 $J$, 0.04 $J$
 and 0.04 $J$.

\noindent {9.} Low-lying energy levels of two spin-1 sawtooth chains (in units
of $J$). The triplet gaps are equal to 0.25 $J$ and 0.24 $J$.
The singlet gaps are equal to 0.5 $J$ and 0.5 $J$.

\newpage

\begin{figure}[ht]
\begin{center}
\hspace*{-2cm}
\epsfig{figure=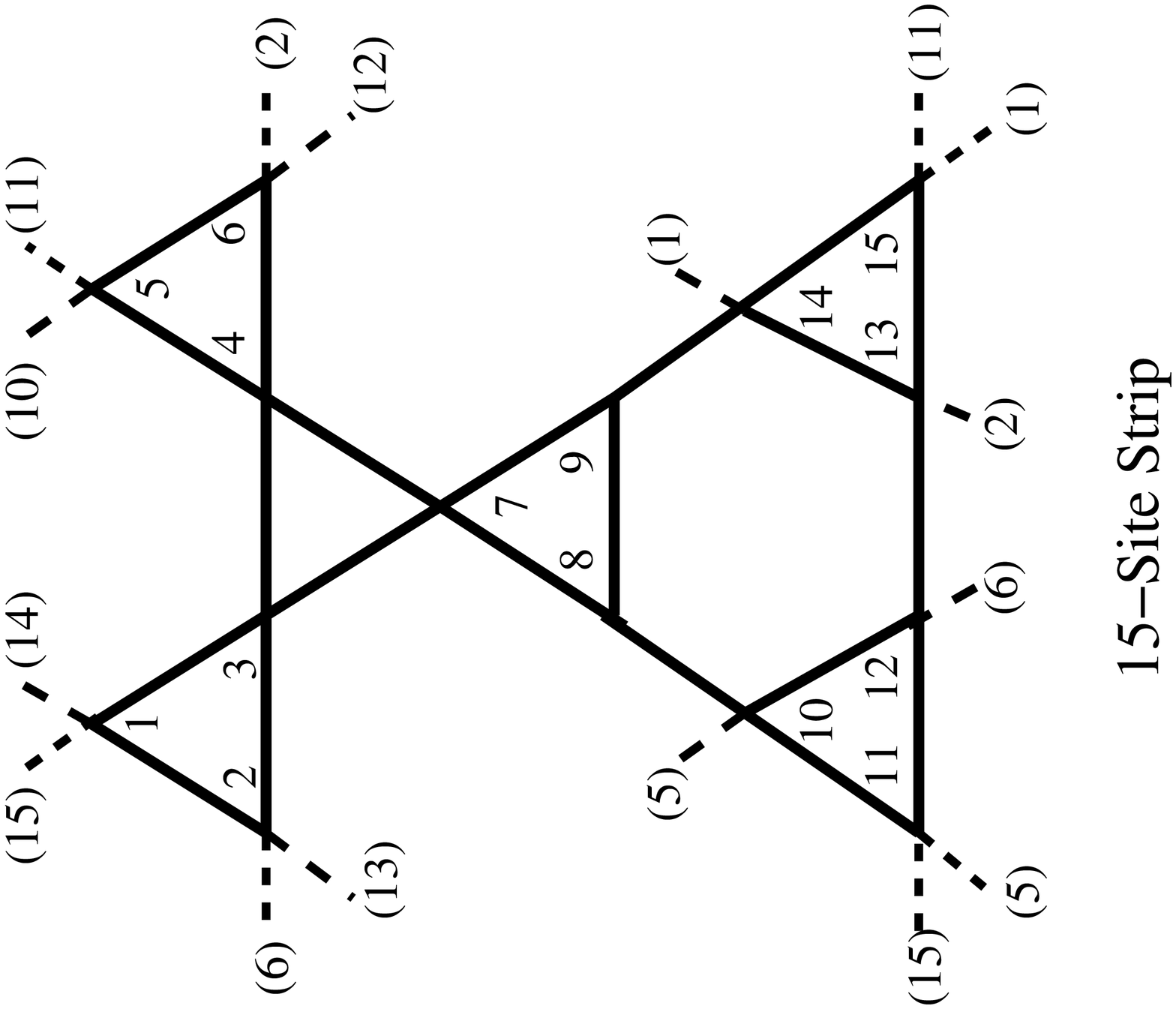,width=18cm,angle=270}
\end{center}
\vspace*{2cm}
\centerline{Fig. 1}
\end{figure}

\newpage

\begin{figure}[hp]
\begin{center}
\hspace*{-2cm}
\epsfig{figure=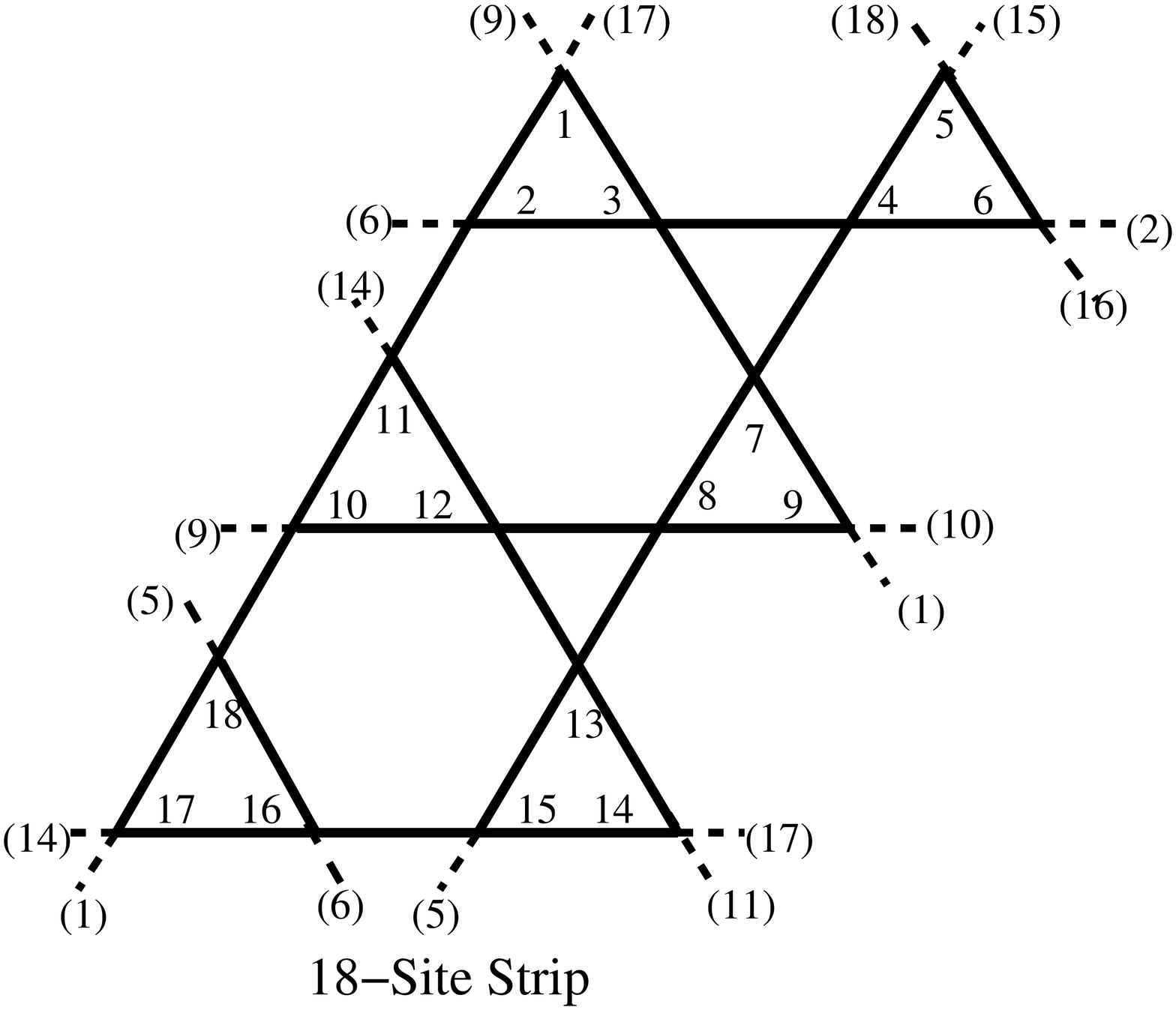,width=18cm,angle=270}
\end{center}
\vspace*{2cm}
\centerline{Fig. 2}
\end{figure}

\newpage

\begin{figure}[hp]
\begin{center}
\hspace*{-2cm}
\epsfig{figure=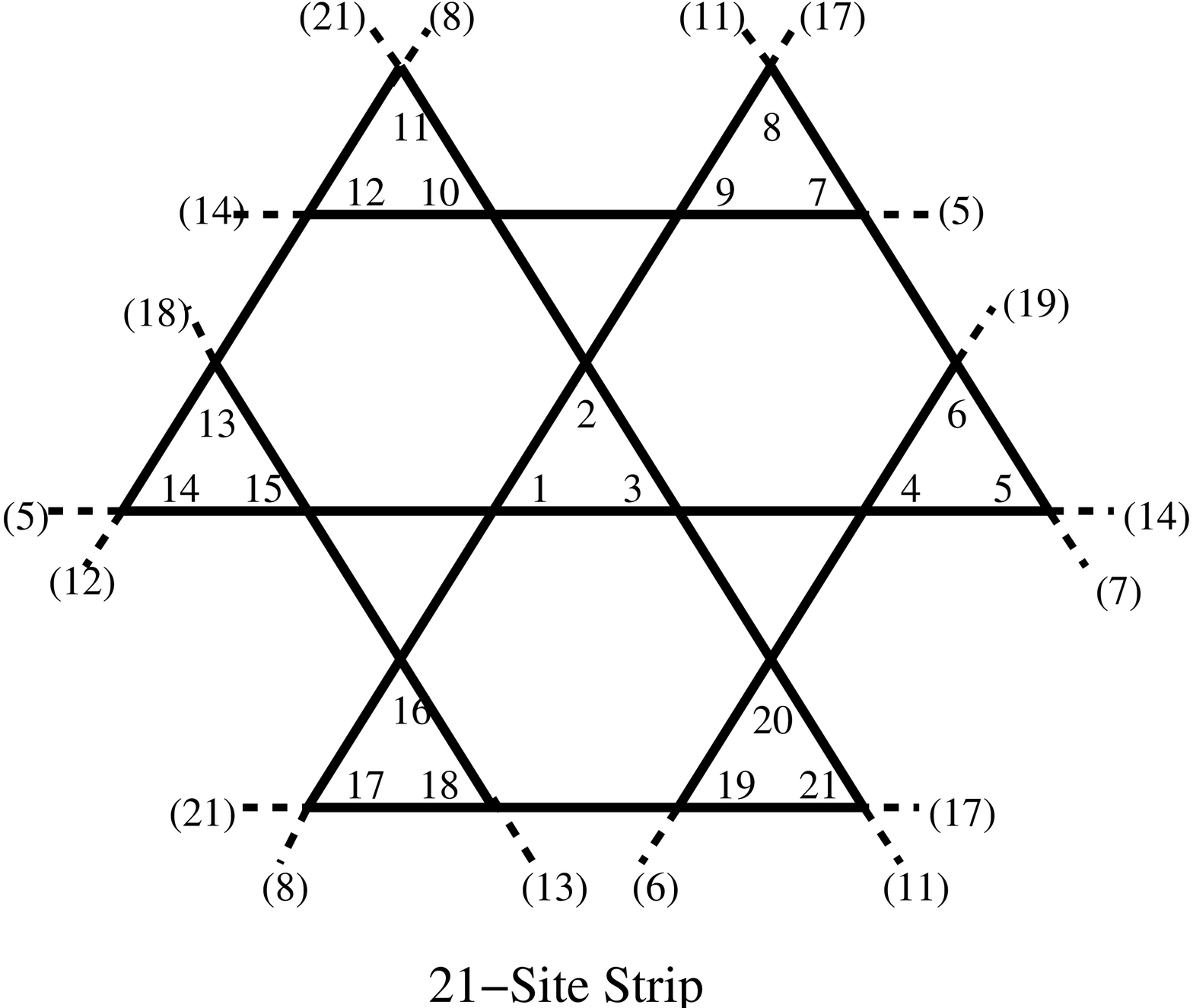,width=18cm,angle=270}
\end{center}
\vspace*{2cm}
\centerline{Fig. 3}
\end{figure}

\newpage

\begin{figure}[hp]
\begin{center}
\hspace*{-2cm}
\epsfig{figure=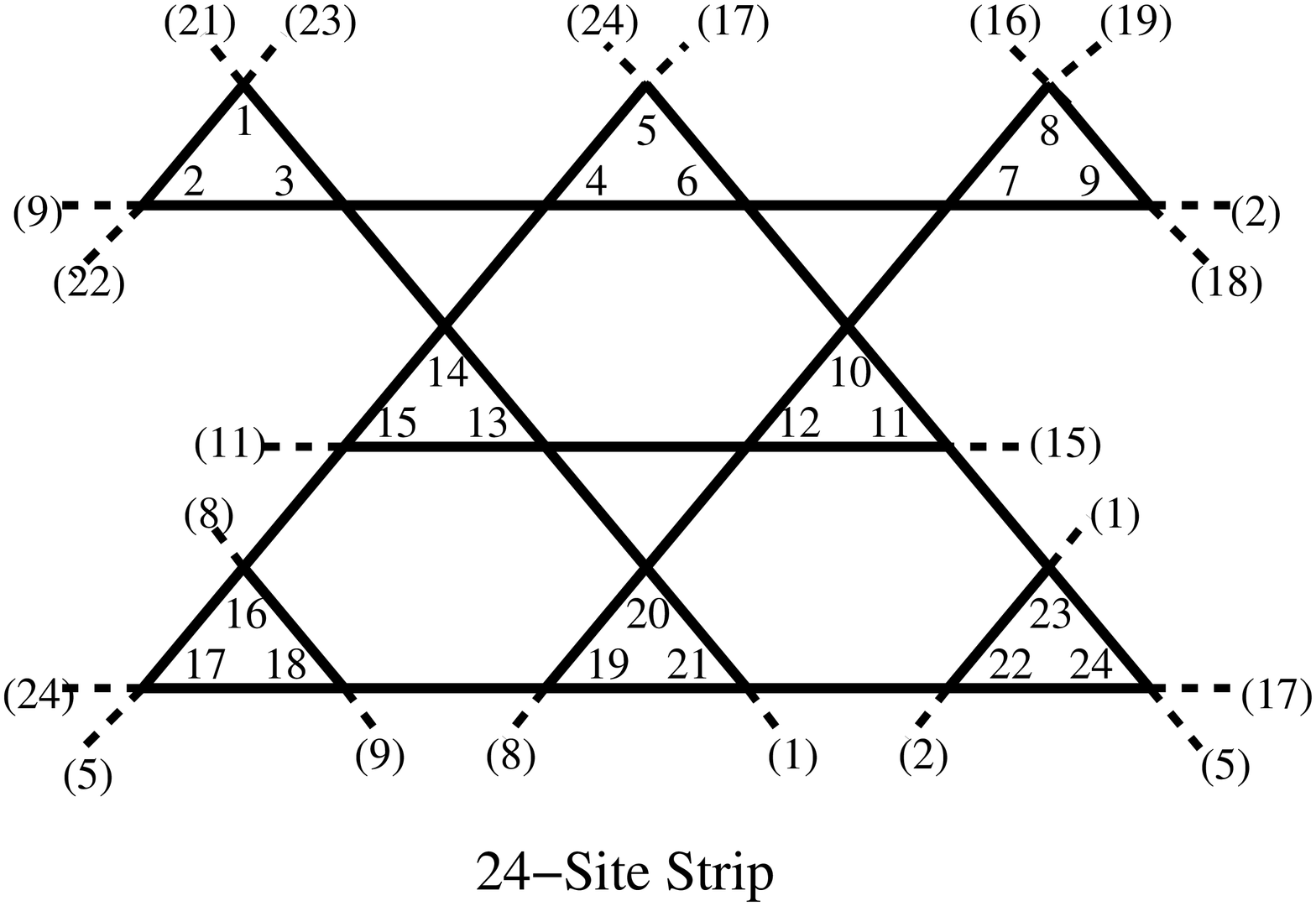,width=18cm,angle=270}
\end{center}
\vspace*{2cm}
\centerline{Fig. 4}
\end{figure}

\newpage

\begin{figure}[hp]
\begin{center}
\hspace*{-2cm}
\epsfig{figure=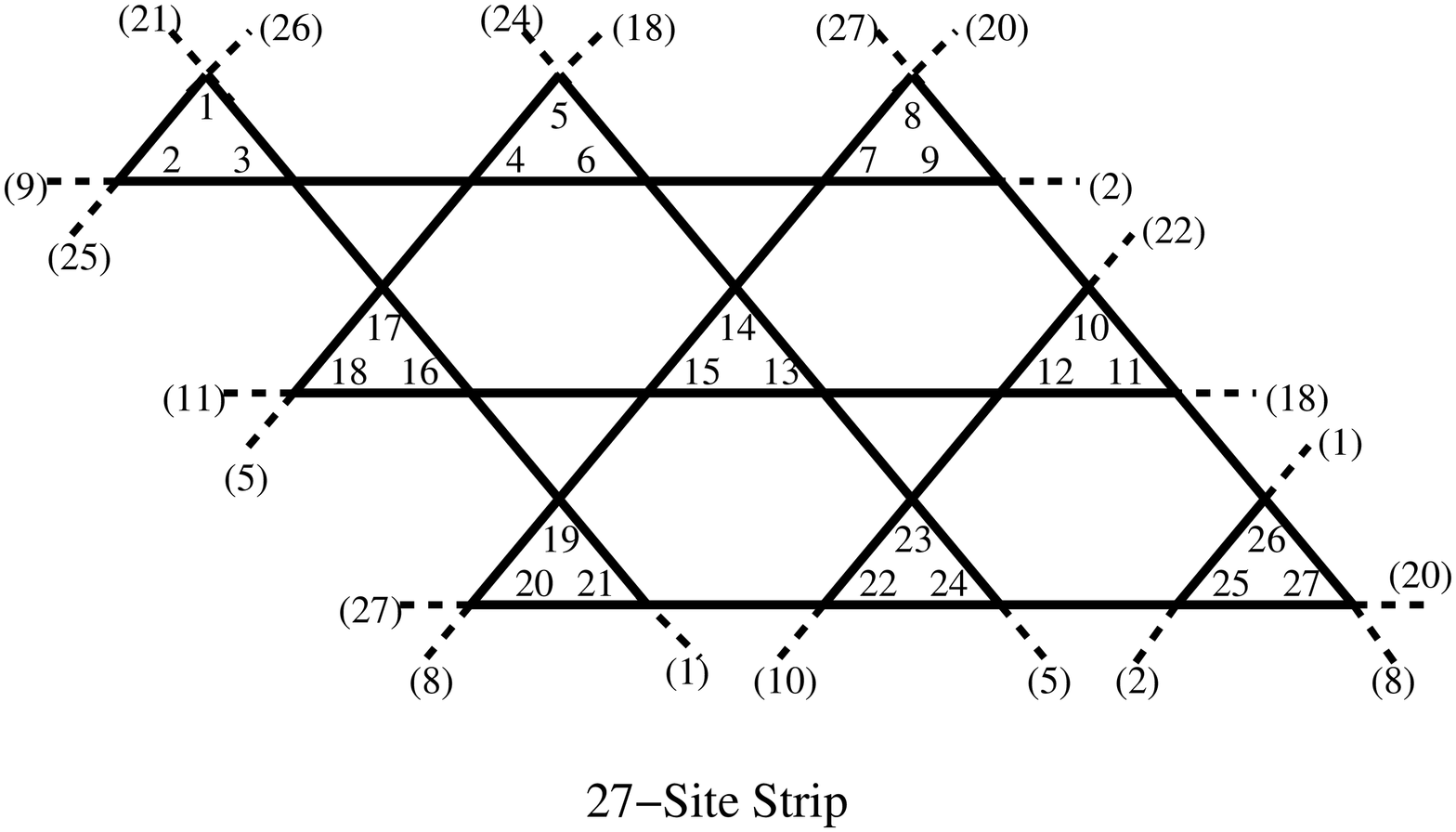,width=20cm,angle=270}
\end{center}
\vspace*{2cm}
\centerline{Fig. 5}
\end{figure}

\newpage

\begin{figure}[hp]
\begin{center}
\hspace*{-1cm}
\epsfig{figure=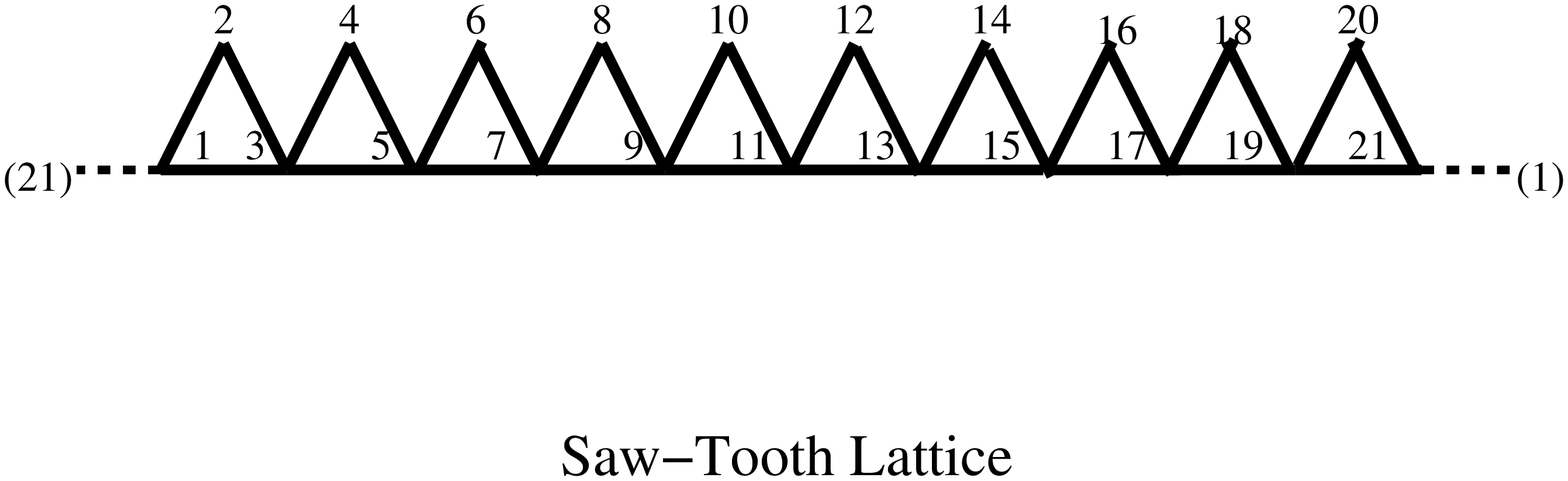,width=19cm,angle=270}
\end{center}
\centerline{Fig. 6}
\end{figure}

\newpage

\begin{figure}[hp]
\begin{center}
\hspace*{-1cm}
\epsfig{figure=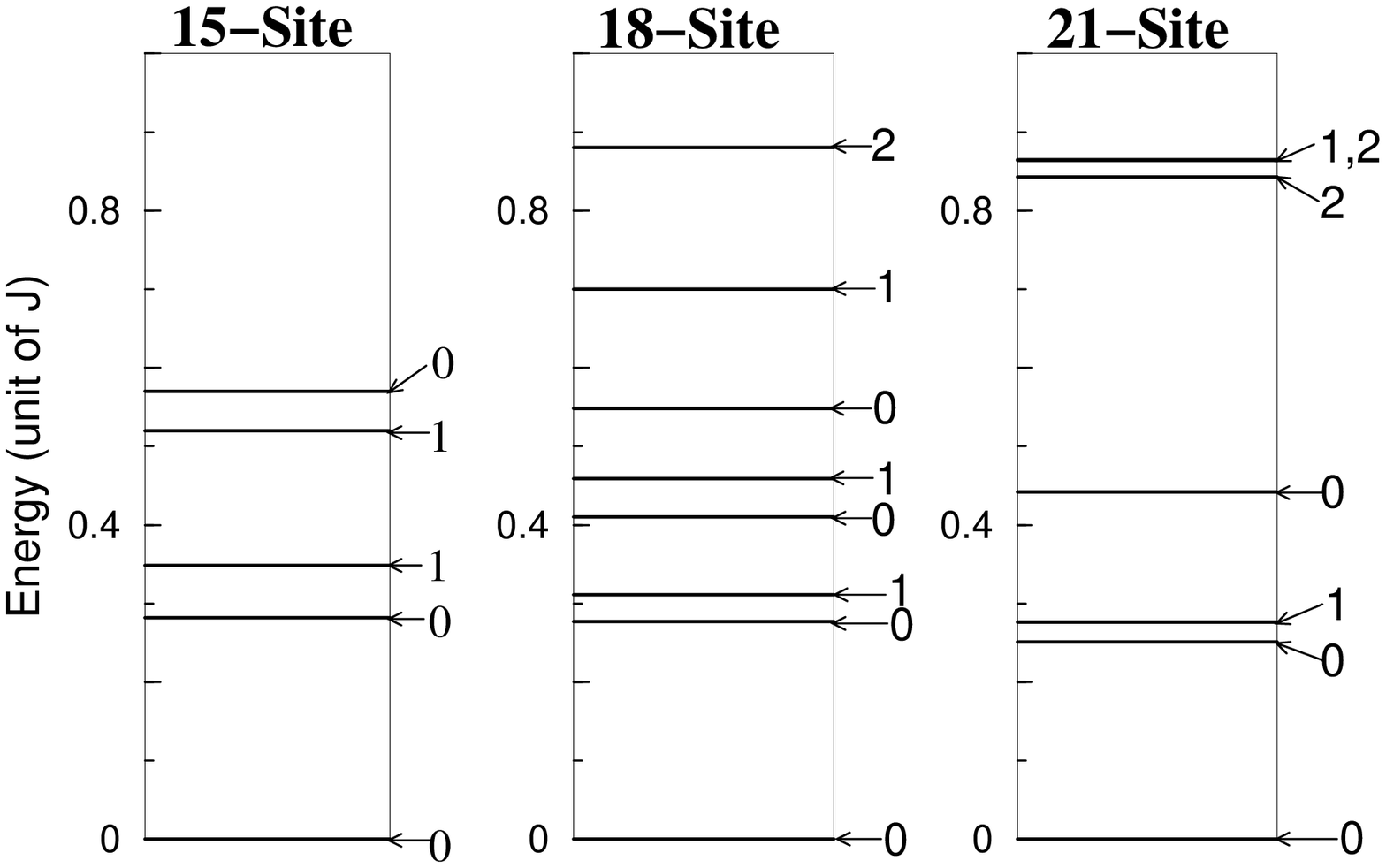,width=16cm}
\end{center}
\vspace*{3cm}
\centerline{Fig. 7}
\end{figure}

\newpage

\begin{figure}[hp]
\begin{center}
\hspace*{-1cm}
\epsfig{figure=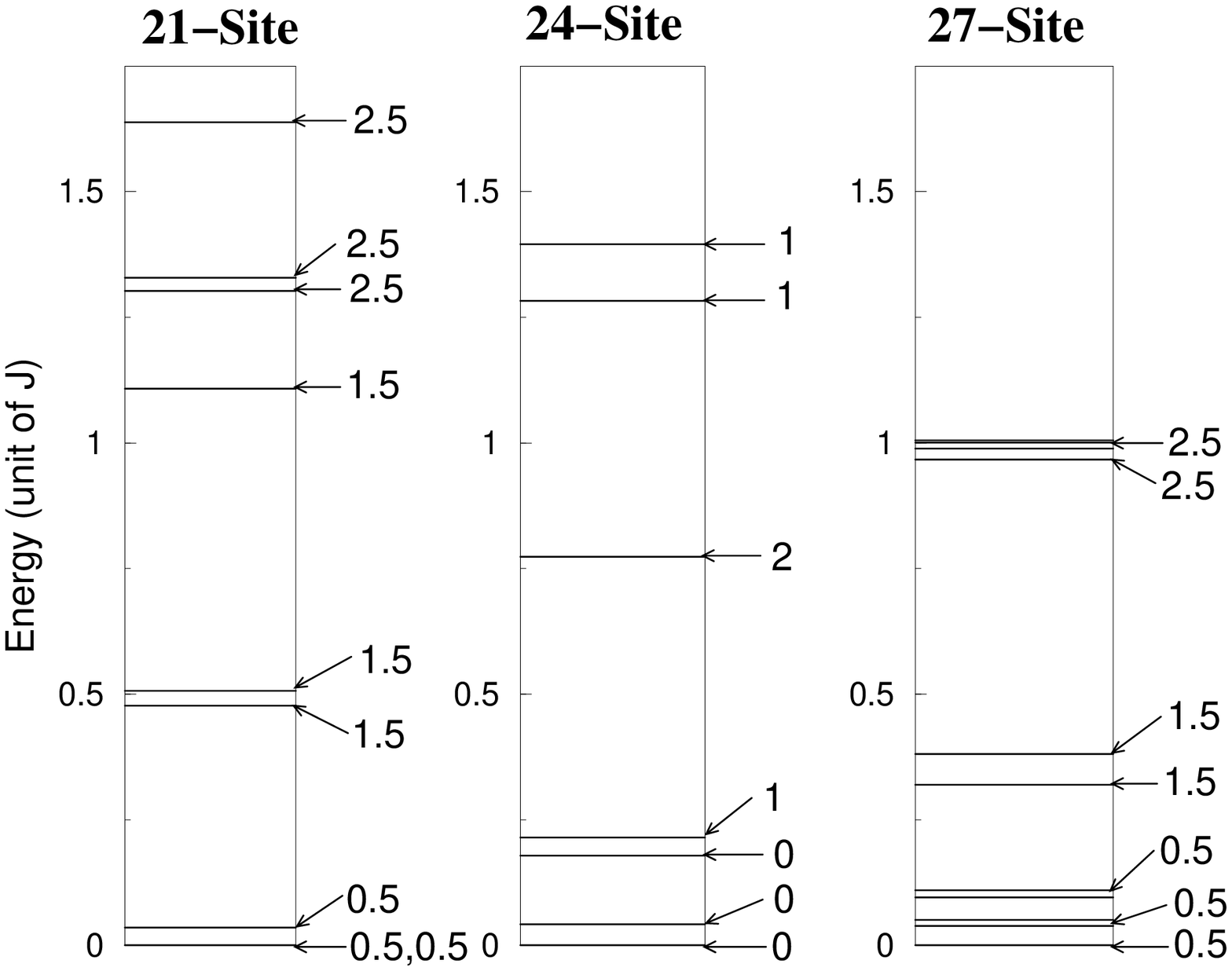,width=16cm}
\end{center}
\vspace*{3cm}
\centerline{Fig. 8}
\end{figure}

\newpage

\begin{figure}[hp]
\begin{center}
\hspace*{-1cm}
\epsfig{figure=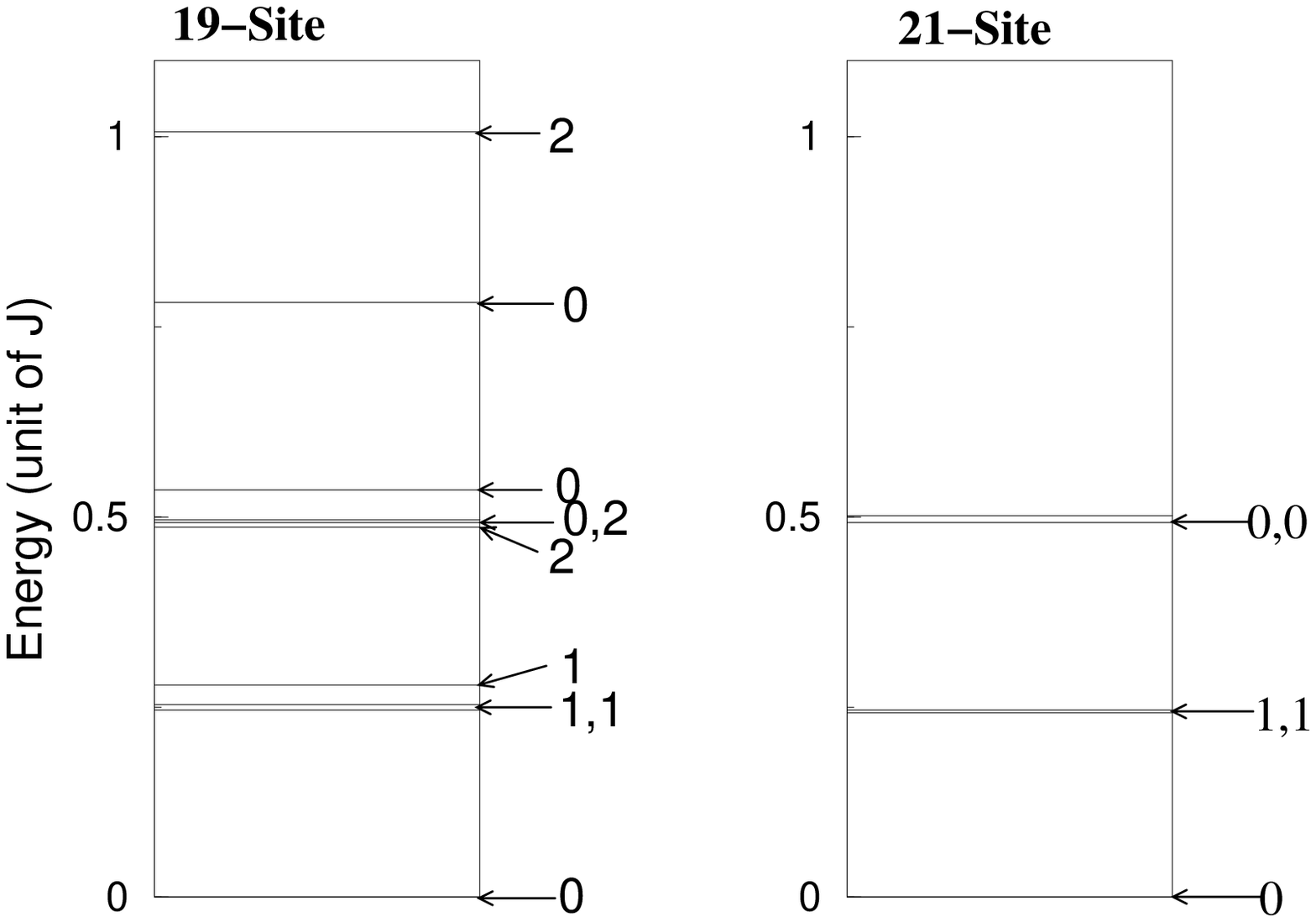,width=16cm}
\end{center}
\vspace*{3cm}
\centerline{Fig. 9}
\end{figure}

\end{document}